\documentclass[journal=enfuem,manuscript=article]{achemso}

\usepackage[version=3]{mhchem} % Formula subscripts using \ce{}

\usepackage{graphicx}
\usepackage{stfloats}
\usepackage{color}
\usepackage{soul}
\usepackage{blindtext}
\usepackage{latexsym,amsmath,amssymb}
\usepackage[T1]{fontenc}
\usepackage[utf8]{inputenc}
\usepackage[english]{babel} 
\usepackage{csquotes}
\usepackage{achemso}

\usepackage{nameref}
\usepackage{varioref}

\usepackage{algorithm2e}
\usepackage{mwe,tikz}
\usepackage[percent]{overpic}
\usepackage{siunitx}
\DeclareSIUnit\atm{atm}
\DeclareSIUnit\bar{bar}
\DeclareSIUnit{\calorie}{cal}
\usepackage{tabularx}
\usepackage[export]{adjustbox}
\usepackage{wrapfig}
\graphicspath{{Figures/}}
\usetikzlibrary{decorations.pathmorphing}
\tikzset{snake it/.style={decorate, decoration=snake}}

\usepackage{booktabs,multirow,multicol}

\usepackage[scriptsize]{subfigure}

\bibpunct{{\unskip~[}}{]}{,}{n}{}{;} % put the ~ in "~[1]" so you don't have to. First removes a space if there is one there.
\usepackage{threeparttable}

\usepackage[breaklinks=true, linkcolor=blue, citecolor=blue, colorlinks=true]{hyperref}
\usepackage[retainorgcmds]{IEEEtrantools}

\usepackage{array}
\usepackage{cleveref}
\usepackage{comment}
\author{Aaron J. Fillo}
\author{Jonathan Bonebrake}
\author{David L.~Blunck}
\email{David.Blunck@oregonstate.edu}
\affiliation[Oregon State Universtiy]
{School of Mechanical, Industrial, and Manufacturing Engineering, Oregon State University, Corvallis, OR 97331, USA}

\title{ Impact of fuel chemistry on the global consumption speed of large hydrocarbon fuel/air flames }

\keywords{Turbulent flames\sep Turbulent flame speed\sep Large hydrocarbon\sep Flame stretch\sep Bunsen flame\sep Turbulent consumption speed}

\begin{document}

% \begin{tocentry}

% Some journals require a graphical entry for the Table of Contents.
% This should be laid out ``print ready'' so that the sizing of the

% Inside the \texttt{tocentry} environment, the font used is Helvetica
% 8\,pt, as required by \emph{Journal of the American Chemical

% The surrounding frame is 9\,cm by 3.5\,cm, which is the maximum
% permitted for  \emph{Journal of the American Chemical Society}
% graphical table of content entries. The box will not resize if the
% content is too big: instead it will overflow the edge of the box.

% This box and the associated title will always be printed on a
% separate page at the end of the document.

% \end{tocentry}

\begin{abstract} 
Large hydrocarbon fuels are used for ground and air transportation and will be for the foreseeable future.
Despite their extensive use, turbulent combustion of large hydrocarbon fuels, such as jet fuels, remains relatively poorly understood and difficult to predict.
A key parameter when burning these fuels is the turbulent consumption speed, which is the velocity at which fuel and air are consumed through a turbulent flame front.
Such information can be useful as a model input parameter and for validation of modeled results.
In this study, turbulent consumption speeds were measured for three jet-like fuels using a premixed turbulent Bunsen burner.
The burner was used to independently control turbulence intensity, unburned temperature, and equivalence ratio.
Each fuel had similar heat releases (within \SI{2}{\percent}), laminar flame speeds (within \SIrange{5}{15}{\percent}), and adiabatic flame temperatures.
Despite this similarity, For constant $Re_D$ and turbulence intensity, A2 (i.e., jet-A) has the highest turbulent flame speeds and remains stable (i.e., without tip quenching) at lower $\phi$ than the other fuels evaluated.
In contrast the C1 fuel, which contains no aromatics, has the slowest turbulent flame speeds and exhibits tip quenching at higher $\phi$ then the other fuels.
C1 was the most sensitive to the influence of turbulence, as evidenced by this fuel having the largest ratio of turbulent to laminar flame speeds.
The C1 fuel had the highest stretch sensitivity, in general, as indicated by calculated Markstein numbers.
This work shows that turbulent flame speeds and tip stability of multi-component large hydrocarbon fuels can be sensitive to the chemical class of its components.
The results from the current work indicate that caution may be needed when using alternative or surrogate fuels to replicate conventional fuels, especially if the alternative fuels are missing chemical classes of fuels that influence stretch sensitivities.
\end{abstract}


%%%%%%%%%%%%%%%%%%%%%%%%%%%%%%%%%%%%%%%%%%%%%%%%%%%%%%%%%%%%%%%%%%%%

\section{Introduction}
Reactions in internal combustion and gas turbine engines operate at elevated temperatures and pressures and are primarily controlled by turbulent premixed, and partially premixed combustion \cite{Law2006CombustionPhysics}.
These applications generally use large hydrocarbon fuels such as gasoline, diesel, and jet fuel because of their high energy-density.
However, turbulent flame speeds for most large hydrocarbon fuels are not known.
Moreover, results from Venkateswaren et al.~\cite{Venkateswaran2013} and Won et al.~\cite{Won2014} have indicated that measured turbulent flame speeds may be as much as \SI{50}{\percent} different from those expected in models.
As such, an improved understanding of turbulent combustion of large hydrocarbon fuels is warranted to support engine development and alternative fuel research.

The turbulent flame speed is a metric for assessing the effects of turbulent fluctuations, molecular transport, and fuel chemistry on turbulent combustion \cite{Won2014}. Four definitions of the turbulent flame speed have been used in literature: local displacement speed ($S_{T,LD}$), global displacement speed ($S_{T,GD}$), local consumption speed ($S_{T,LC}$), and global consumption speed ($S_{T,GC}$) \cite{Venkateswaran2011,GouldinF.andCheng2010,Cheng2009,DRISCOLL2008,Verma2016}.
Displacement and consumption refer to the propagation of the flame front and consumption of reactants (mass burning flux), respectively.
Local displacement and consumption speeds are challenging to obtain experimentally because the relevant time and length scales are typically on the order of Taylor or Kolmogorov length-scales, requiring more advanced diagnostics to capture~\cite{Borghi1985OnFlames}.

% For reference the definition for the local consumption speed is
% \begin{equation} \label{eq2.4}
% S_{T,LC} = S_{L,0}I_{0}\int_{-\infty}^{\infty}\Sigma d\eta\;, %variable definitions in Venkateswaran 2011
% \end{equation}
% where $I_{0}$ and $\Sigma$ are the stretch factor and the flamelet surface area per unit volume, respectively \cite{Venkateswaran2011}.
% The surface area is integrated along the path of $\eta$, over the flame brush, in the direction normal to the reaction surface  \cite{Law2006,Hemchandra2010}. 
% This approach is experimentally inaccessible and analytically intensive making it difficult to implement.
% As demonstrated by Cheng et al.\ \cite{Cheng2008} the local displacement speed is similarly intensive in its application.
Measurements of global burning velocities can be used to provide insights into the impact of turbulence on combustion behavior while avoiding challenges associated with local flame speed measurements \cite{Law2006,Venkateswaran2011}.
The turbulent consumption speed in particular represents a temporally and spatially averaged measure of reactant consumption through the flame brush, defined as,
\begin{equation} \label{eq2.5}
S_{T,GC}=\frac{\dot{m}_{R}}{\rho_{R}\overline{A}_{\langle c\rangle}}\;,
\end{equation}
where $\dot{m}_{R}$, $\rho_{R}$, and $\overline{A}_{\langle c\rangle}$ are the mass flow rate of the reactants, the density of the reactants, and the mean flame area corresponding to the contour $\langle c\rangle$ \cite{Venkateswaran2011, GouldinF.andCheng2010,Cheng2009}.
The contour $\langle c\rangle$ corresponds to the progress variable, $c$, which is a measure of reactant consumption through the flame brush.
Here, $\langle c\rangle=0$ and $\langle c\rangle=1$ are the unburned and burned faces of the flame brush, respectively~\cite{Kobayashi1996}.

Studies of the turbulent consumption speed for small hydrocarbon fuels (e.g., gaseous at room temperature and pressure) have provided insights into turbulence and chemistry interactions that can occur in engines \cite{GouldinF.andCheng2010,Venkateswaran2011,Venkateswaran2013,Kobayashi1996,Kobayashi1998,Kobayashi2002}.
For example, work by Venkateswaran et al.\ \cite{Venkateswaran2011} measured the turbulent consumption speeds of premixed \ce{CH_{4}} and \ce{H_{2}/CO} (Syngas) fuel blends of \SI{30}{\percent}, \SI{50}{\percent}, \SI{70}{\percent}, and \SI{90}{\percent} \ce{H_{2}} by volume at atmospheric conditions on a piloted turbulent Bunsen burner \cite{Venkateswaran2011}.
These mixtures had matched zero-stretch laminar flame speeds but different Lewis numbers.
The turbulent consumption speed increased with increasing turbulence intensity and increased proportionally to the percent weight of \ce{H_{2}}.
Methane/air flames had the lowest observed flame speeds.
This sensitivity of the turbulent consumption speed to the fuel composition, for fuels with similar laminar flame speeds, was attributed to preferential diffusion effects.
Moreover, this study observed that the turbulent flame speeds for \SI{90}{\percent} \ce{H_{2}} fuel blends were three times larger than methane/air flames with the same laminar flame speeds.
% This indicates that classic correlations between laminar flame speed and turbulence intensity (non-dimensional velocity fluctuations, $I=u_{rms}'/U_{0}$) can be insufficient for correlating the turbulent flame speed.
In subsequent research, Venkateswaran et al. \cite{Venkateswaran2013} measured the turbulent flame speeds of \ce{H_{2}/CO} fuel blends at pressures of \SIrange{1}{20}{atm}.
Similar to previous studies, different fuels with equivalent unstretched laminar flame speeds had different turbulent flame speeds (up to 60\%) even for the same turbulence intensities.
Venkateswaren et al. \cite{Venkateswaran2015, Venkateswaran2015a} attributed the differences in flame speeds, in part, to a coupling between turbulence, chemistry, and pressure effects.
They further attributed the differences to a sensitivity of the turbulent consumption speed to flame stretch rate.  
What remains to be determined is if the turbulent flame speed of large hydrocarbon fuels, such as jet fuels, are similar for fuels with comparable laminar flame speeds.

% \subsection{Flame Stretch Sensitivities}

A sensitivity of flame stretch to fuel chemistry has been observed for multi-component fuel blends used for aviation.
Kumar et al.~\cite{Kumar2011} investigated the laminar flame speeds and stretch-based extinction limits of three conventional and alternative fuels including Jet-A, S-8 (a synthetic jet-fuel), and pure n-decane and n-dodecane.
Jet-A and S-8 demonstrated similar propagation characteristics but notable differences in extinction limits.
S-8 flames remained stable at stretch rates \SI{10}{\percent} or more greater than Jet-A flames at similar conditions.
Both pure n-decane and n-dodecane  demonstrated the greatest stability and remained stable at stretch rates as much as \SI{20}{\percent} greater than Jet-A and S-8 at similar conditions.
This increased extinction stretch rate was attributed to the reduced aromatic content; of the four fuels studied, only Jet-A contained aromatics.
What was not considered in this study was the potential influence of normal- or iso- structure of the fuels.

Turbulent Bunsen flames exhibit both hydrodynamic (local flame wrinkling) and curvature-based (global flame curvature) stretch effects.
These effects are heavily coupled and do not lend themselves well to analytical solutions, as is possible with laminar Bunsen flames.
However, these principles of hydrodynamic and curvature effects can be applied to turbulent Bunsen flames, such as those used in this study, through application of dimensional analysis.
Specifically, parameters such as the root mean square of turbulent fluctuations ($u'_{rms}$) or the unburned bulk flow velocity ($U_o$) can be scaled by the unstretched laminar flame speed ($S_{L,0}$), highlighting hydrodynamic or curvature-based stretch effects, respectively.

The limited data reported for the turbulent flame speeds of large hydrocarbon fuels (i.e., liquid at room temperatures) has focused on the flame speeds for single component fuels.
Moreover, the flame speeds were collected on different burners for different fuels, making it challenging to compare values for different fuels.
Goh et al.\ \cite{Goh2012} measured turbulent consumption speeds for JP-10/air mixtures at \SI{473}{\kelvin} and atmospheric pressure for a range of $\phi=$ \num{0.2} to \num{0.8} on a twin-flame opposed-flow burner.
They observed an increase in turbulent flame velocities with increasing axial root mean square (rms) velocity fluctuations and Damk\"ohler number.
These results suggest turbulent flame speed is sensitive to turbulent time and length scales. 
However, the published range of data is small, and further investigation is needed to confirm apparent sensitives of turbulent consumption speeds. 
In another study investigating large hydrocarbon fuels, Won et al.\ \cite{Won2014} measured the turbulent burning velocities of n-heptane/air mixtures using a Reactor Assisted Turbulent Slot (RATS) burner.
The study demonstrated that the turbulent flame speed is sensitive to low-temperature chemistry by varying the preheat temperature while holding flow velocity constant.
The study also showed that the turbulent flame speed is sensitive to pre-flame oxidation which altered fuel chemistry and transport properties.
This result verifies that the turbulent flame speed is sensitive to fuel chemistry.% since they are independent of geometry and flame speed definition, as the low temperature effects are determined upstream of the burner.

In summary, sensitivities of the global consumption speed to turbulence intensity, pressure, preferential diffusion, and fuel chemistry have been well documented \cite{Won2014,Venkateswaran2011,Venkateswaran2015,Daniele2012,Kobayashi1998} for small hydrocarbon fuels, but much less is known about these sensitivities for large hydrocarbon fuels, in particular multi-component fuels.
With this background, the objective of this effort is to identify sensitivities of the turbulent consumption speed of multi-component large hydrocarbon liquid fuels to fuel chemistry, bulk Reynolds number, and turbulence intensity.  
Jet-A and two large hydrocarbon fuels with similar heat release rates and laminar flame speeds, but different chemical compositions, are evaluated.  This work adds to the sparse set of fundamental data describing turbulent premixed combustion of large hydrocarbon fuels representative of those used for aviation.  The turbulent flame speeds reported in this work can be used as input parameters for some models of reacting flows and can be used to help evaluate the fidelity of modeled results of burning large hydrocarbon fuels \cite{Khalilarya2010, Briones}.  Moreover, potential chemical sensitivities to flame tip quenching are identified.

\section{Experimental approach and facility}
\begin{figure}[!ht] 
  \begin{center}
  \includegraphics[width=0.6\textwidth]{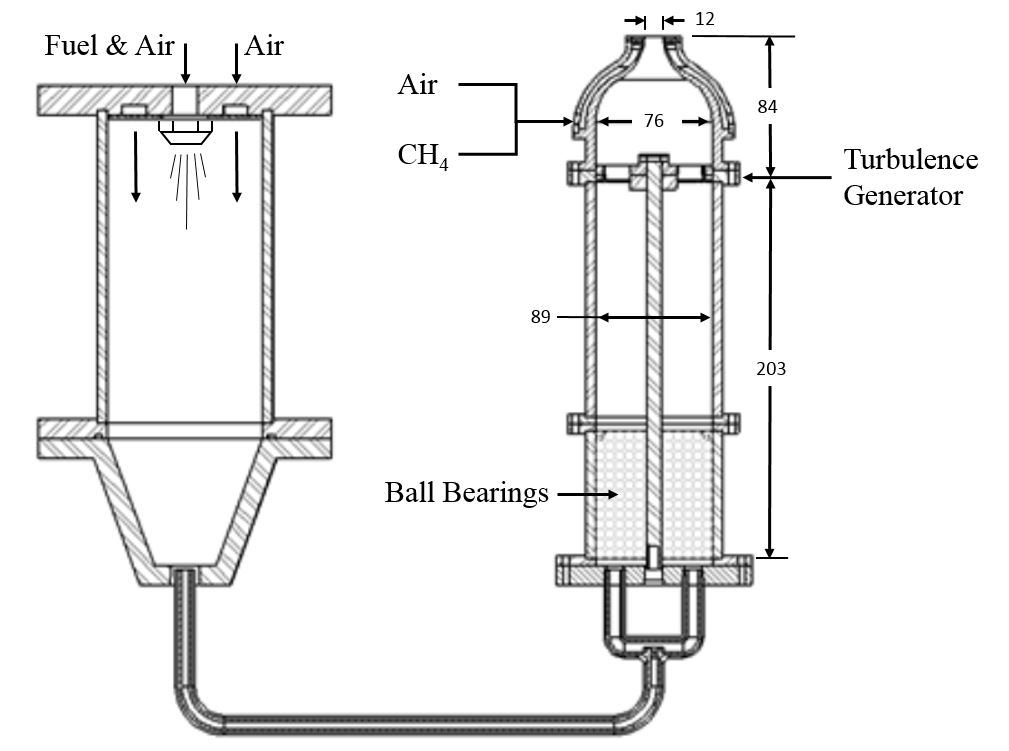}
  \caption{Schematic of the vaporizer and burner system.  All dimensions are in mm.}
  \label{fig:schematic}
  \end{center}
\end{figure}

% \subsection{Experimental arrangement}
A schematic of the burner system used in this study is presented in Figure \ref{fig:schematic}. 
This burner replicated the design developed by Venkateswaran and colleagues \cite{Venkateswaran2011}.
The system was used to independently control the Reynolds number, turbulence intensity, and unburned temperature of the flow.
%The burner consisted of a smooth contoured nozzle tapering from \SI{76}{\milli\meter} at the inlet to an exit diameter of \SI{12}{\milli\meter}.
The major components included the vaporizer, mixing section, burner plenum, turbulence generator, and burner nozzle.

In the vaporizer, preheated air ($\approx$ \SI{470}{\kelvin}) and room temperature liquid fuel were injected into a cylindrical chamber using an air assisted atomizing nozzle.
Additional preheated air was injected into the vaporizer through an annular manifold positioned around the fuel nozzle.
The manifold maintained a high velocity heated air-curtain around the fuel injection site, encouraging turbulent mixing and helping to prevent fuel from contacting the heated vaporizer walls.
Downstream of the vaporizer, the fuel-air mixture was maintained at a constant temperature of $\approx$ \SI{470}{\kelvin} prior to entering the burner plenum.

In the plenum, the jet-A/air mixture passed through a layer of ball bearings upstream of the nozzle; this prevented “jetting” from the smaller diameter feed-lines and encouraged uniform flow development \cite{Venkateswaran2011}.
The flow then passed through a short development length to ensure it was well mixed with a uniform temperature and velocity distribution.
A type-K thermocouple was used to monitor the final unburned temperature of the reactant flow ($\approx$ \SI{450}{\kelvin}) near the outlet.  
%The manufacturer specified uncertainty of the temperature measurement was \SI{2}{\percent} of the measured value.
Care was taken to keep the mixture temperature below auto-ignition temperatures at all times for safety and to minimize any fuel cracking that might occur \cite{Won2014}.

% Vaporization of the liquid fuels, and mixing with the pre-heated air stream, was performed in a vaporizer located upstream of the burner. 
%Vaporization of the liquid fuel and mixing with preheated air was performed in a vaporizer upstream of the burner assembly.
%The vaporizer section premixes and vaporizes the liquid fuels.
%This component replicates a system developed by Air Force Research Laboratory.
% Preheated air ($\approx$ \SI{470}{\kelvin}) and room temperature fuel were injected into the vaporizer using an air assisted atomizing nozzle.
% Additional preheated air was injected into the vaporizer through an annular manifold positioned around the fuel nozzle.
% The manifold maintained a high velocity heated air-curtain around the fuel injection site, encouraging turbulent mixing and helping to prevent fuel from contacting the heated vaporizer walls.
% Downstream of the vaporizer, the reactant fuel-air mixture was maintained at a constant temperature of $\approx$ \SI{470}{\kelvin} prior to entering the burner plenum. The vaporizer was located $\approx$ \SI{1.3}{\meter} upstream of the burner outlet, helping to ensure that a well-mixed reactant flow entered the burner nozzle.
% Care was taken to keep the mixture temperature below auto-ignition temperatures at all times for safety and to minimize any fuel cracking that might occur \cite{Won2014}.

The  turbulence  generator  was located 84 mm upstream of the burner outlet. 
The device consisted of a fixed lower plate and rotating upper plate with radial openings permitting flow through the assembly.
Rotation of the top plate relative to the bottom plate
changed the blockage ratio of the turbulence generator and consequently the turbulence intensity. 
Flow straighteners attached to the top and bottom plates reduced the degree of swirl initiated by the turbulence generator.
The turbulence intensity of the flow was proportional to the blockage ratio, which was determined using a high contrast top down photograph of the upper plate position.
The turbulence intensities were estimated based on the blockage ratio and data reported by Venkateswaran et al. \cite{Venkateswaran2011,Venkateswaran2013} for a matching turbulence generator and burner geometry.
Further detail on the turbulence generator construction can be found in reference \cite{Venkateswaran2011}.

Exiting the turbulence generator, the flow entered the burner nozzle.
The nozzle was designed to reduce boundary layer growth and produce a top hat velocity profile at the exit \cite{Venkateswaran2011}.
An annular sintered plate with a \SI{20}{\micro\meter} nominal pore size was positioned around the burner outlet and anchored the premixed methane/air pilot flame.
The pilot flame was operated at an equivalence ratio ($\phi$) equal to 1 with a heat release rate $\approx$ \SI{10}{\percent} of the heat release rate for the Bunsen flame.

All tests were repeated for a range of equivalence ratios from \numrange{0.7}{1} at a pressure of \SI{1}{atm}.  The air flow rates for both the pilot and main flames were metered upstream of the vaporizer and preheaters using high accuracy rotameters (Omega FL3696ST and FL4611-V, respectively) with $\approx$ \SI{3}{\percent} full-scale accuracy.  Pressure transducers with \SI{2}{\percent} full-scale accuracy and type-K thermocouples were used to correct for density changes within the rotameters.
The mass flow rate of the methane used for the pilot flame was metered directly using a thermal mass flow controller (MKS M100B-2000) with a full-scale accuracy of \SI{1}{\percent}.
The main fuel flowrate was metered volumetrically using dual syringe-pumps (ISCO 100DX) with an accuracy of \SI{0.5}{\percent} of the set point value.
The instrument uncertainty of the equivalence ratios for the pilot and main flows were below \SI{3}{\percent} and \SI{5}{\percent}, respectively.

The Reynolds numbers of the flames evaluated in this study were varied from \num{5000} to \num{10000} with turbulence intensities of $\approx$ \SI{10}{\percent} and $\approx$ \SI{20}{\percent} of the bulk flow velocity.
Venkateswaran et al.~\cite{Venkateswaran2011} and Marshall et al.~\cite{Marshall2012} conducted a thorough characterization of the turbulent statistics of flows through this Bunsen burner and turbulence generator configuration.
Turbulence intensity was found to increase monotonically with blockage ratio; this trend was consistent over a wide range of temperatures, pressures, and bulk flow velocities. The turbulence intensities of the flames reported in this work (i.e., $\approx$ \SI{10}{\percent} and $\approx$ \SI{20}{\percent} of the bulk flow velocity) correspond to the minimum and maximum blockage ratios as characterized by Marshall et al.~\cite{Marshall2012}.
% The uncertainty of the turbulence intensity is assumed to be \SI{5}{\percent} of $u^{'}_{rms}/U_0$, within 95 $\percent$ confidence. 
%These turbulence statistics are assumed to have an uncertainty of \SI{10}{\percent} of the bulk flow velocity.

\begin{figure*}[htb!]
  \centering
  \subfigure[Bulk velocity]{\label{fig:vel_profile-a}\includegraphics[width=0.35\textwidth]{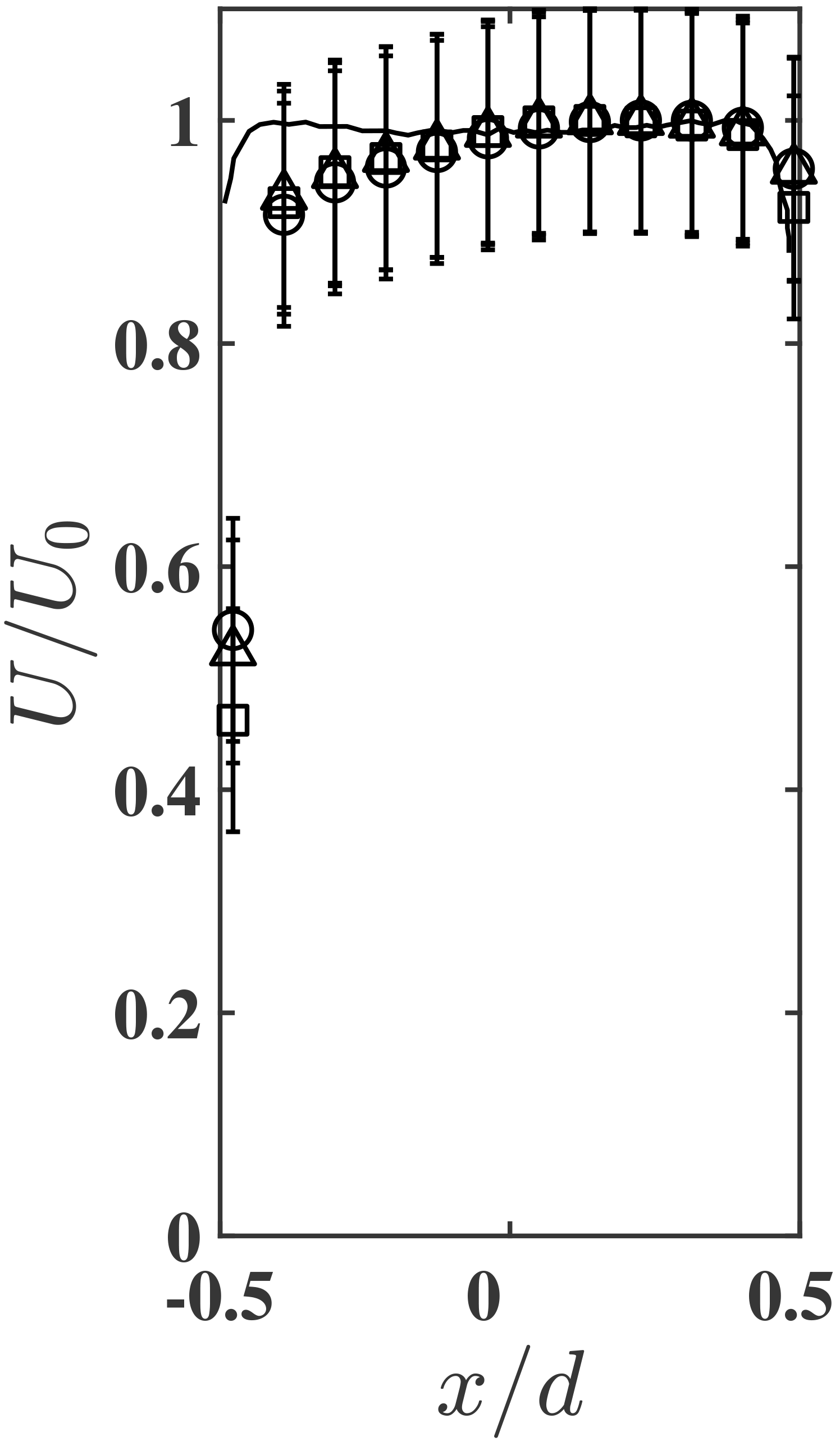}}
  \subfigure[RMS velocity]{\label{fig:vel_profile-b}\includegraphics[width=0.35\textwidth]{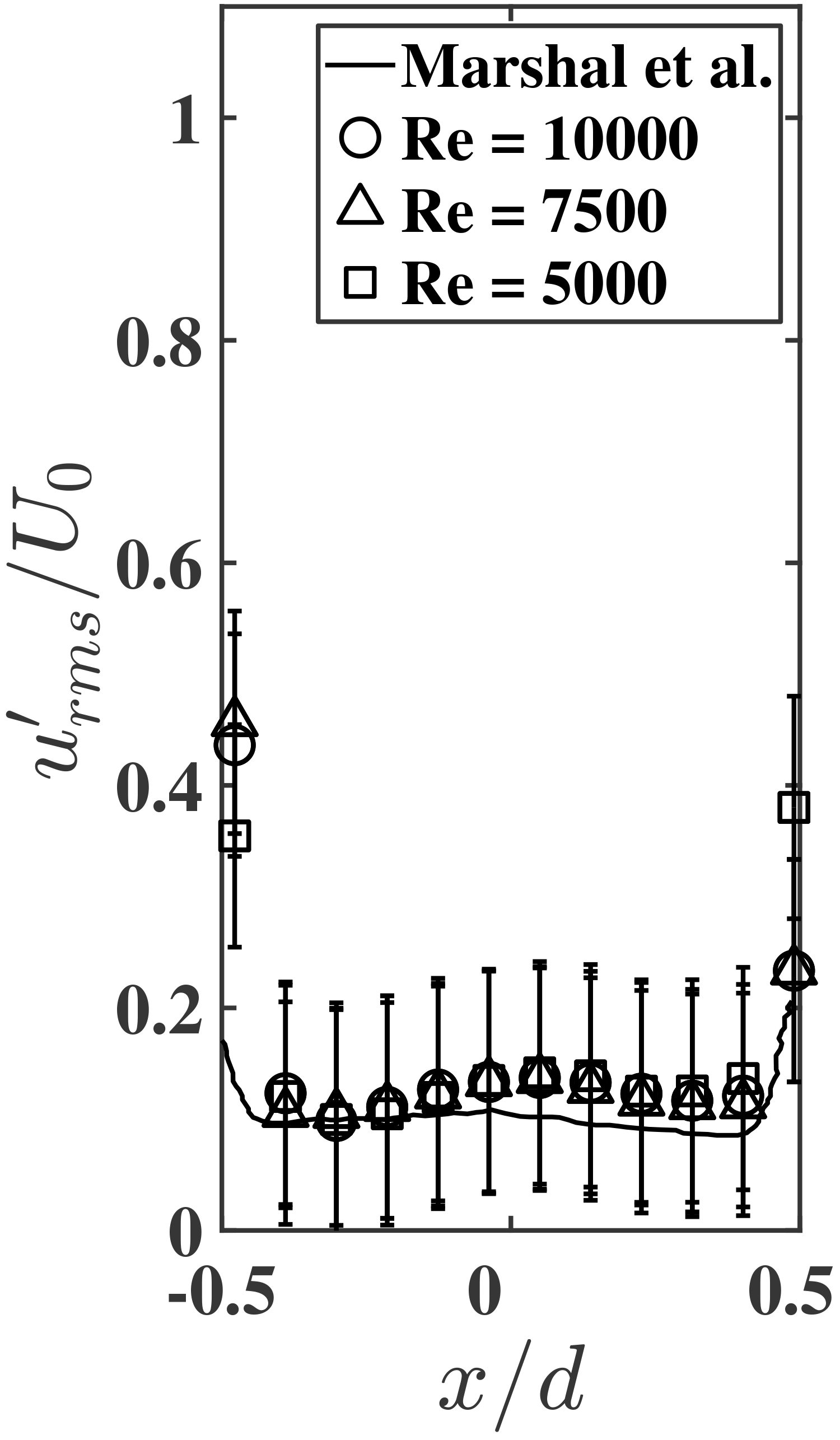}}
  \caption{Normalized axial velocity profiles at 3 mm above the burner exit plane.  The black line represents data reported by Marshall et al. \cite{Marshall2012} and the symbols are the measurements for this study. Panel (a) shows the bulk velocity profile normalized by maximum axial velocity.  Panel (b) shows the distribution of the velocity fluctuations for conditions with turbulence intensity near $\approx$ \SI{10}{\percent}.}
  \label{fig:vel_profile}
\end{figure*}

A brief study was conducted to measure the bulk and fluctuating axial velocity profiles at the burner exit to confirm agreement of results from this study with flow characteristics observed by Marshall et al.~\cite{Venkateswaran2011,Marshall2012}.  A single component hot-wire anemometry probe was used to measure velocities corresponding to Reynolds numbers of \num{5000}, \num{7500}, and \num{10000}. 
The measurements are presented in Figure \ref{fig:vel_profile}, and demonstrate reasonable agreement with results presented by Marshall et al.~\cite{Marshall2012}. 

Finally, the effect of the pilot flame heat release rate on the measured turbulent consumption speed was evaluated. 
Increasing the pilot flame mass flow from \SI{5}{\percent} to \SI{10}{\percent} of the main flame mass flow (which provided the heat release rate of $\approx$ \SI{10}{\percent}) resulted in a \SI{5}{\percent} increase in the turbulent flame speed, which was within the experimental uncertainty.

\subsection{Image processing}
The mean flame brush area ($\bar{A}_{\langle{c}\rangle}$) was determined which in turn was used to determine $S_{T,GC}$. The process has been applied previously \cite{Venkateswaran2011, Venkateswaran2013PressureBlends, Venkateswaran2013} and is now described.
Chemiluminescence measurements of the flame brush were obtained using a \SI{16}{bit} ICCD camera (Andor Solis 334T) with a \num{1024}$\times$\num{1024} pixel resolution and a \SI{25}{\mm}, $f/$\num{4.0}, UV lens. 
The camera was sensitive in the visible and ultraviolet spectrum between \SIrange{230}{1100}{\nano\meter}.
This sensitivity enabled measurement of chemiluminescence from \ce{CH^*}, \ce{OH^*} and \ce{CO2^*}.
The line-of-sight images were obtained over a \SI{3}{\minute} interval at a sampling rate of \SI{2}{\hertz}.
The total sampling time was chosen so that the mean and standard deviation of intensity profiles converged to a constant value before the end of the sampling time.
The exposure time and gate width for the images were \SI{0.1}{\second} and \SI{0.07}{\second}, respectively. 
For each experimental condition, a total of 9 series of 360 line-of-sight images were collected over multiple days. 

\begin{figure*}[htb!]
    \centering
    \includegraphics[width=0.6\textwidth]{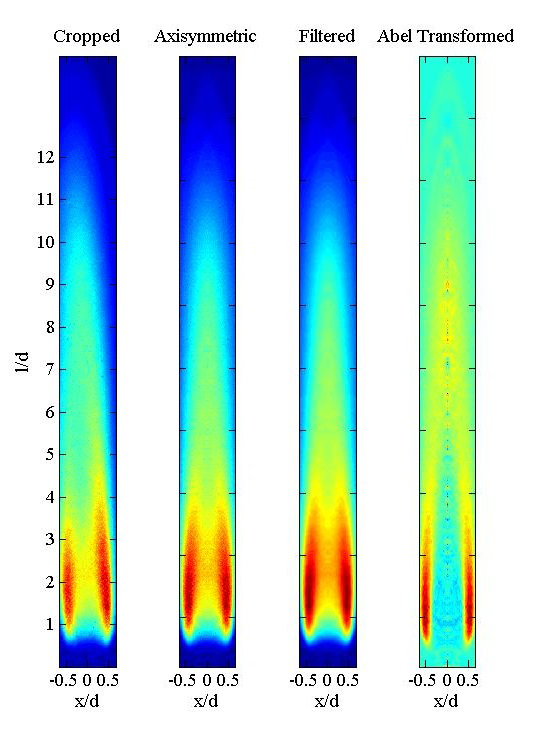}
    \caption{Step-by-step illustration of image processing approach: time-averaged, background subtracted and cropped (panel a), axisymmetric correction (panel b), 2-D median filter (panel c), Abel transform with $\langle{c}\rangle$ = \num{0.5} contour drawn (panel d).}
    \label{fig:img_proc}
\end{figure*}

The average flame sheet location was determined using the image processing technique developed by Venkateswaran et al.~\cite{Venkateswaran2011}.
The steps are summarized in Fig. \ref{fig:img_proc}.
First, the images were time-averaged, the background was subtracted, and the image was cropped to contain only the flame brush.
Next, the image was corrected for any asymmetry and filtered using a 2-D median filter with a kernel less than \SI{2}{\percent} of the burner diameter (5x5 pixels).
A 3 point Abel deconvolution was applied and a Gaussian curve was fit to the resulting centerline intensity profile.
The location of maximum centerline intensity was then determined from the fitted curve.
This location corresponded to the leading edge of the time averaged flame brush.
The leading edge is considered to be the most probable location of the flame brush, and was defined as the $\langle c \rangle = $ \num{0.5} progress variable contour \cite{Venkateswaran2011}.
The sensitivity of flame speeds to the specific progress variable used has been reported previously \cite{Venkateswaran2011}.  Global consumption speeds are inherently approach-specific. What is essential to the validity of the current work, and to the research published previously \cite{Venkateswaran2011, Venkateswaran2013PressureBlends, Venkateswaran2013} is that a consistent progress variable is used.
The estimated uncertainty in the flame speed due to the data reduction process is \SIrange{1}{2}{\percent} \cite{Venkateswaran2011}.
Finally, the location of the leading edge was used to define a conical geometry extending from the location where $\langle c \rangle = $ \num{0.5} to the burner exit plane, corresponding to the apparent average surface area ($A_{\langle c \rangle}$) of the flame brush. % (see Figure \ref{fig:img_proc}(d))

The total bias and precision uncertainty of the measurements are on the order of \SI{10}{\percent}, with a \SI{95}{\percent} confidence interval.  It is noted that Wu et al. \cite{wu2018} conducted a comparison of laminar flame speeds measured using PLIF and OH* chemiluminesence for jet-A and jet-A surrogate fuels.  The difference between values obtained using the two different techniques was 1-3 $\%$ at conditions similar to those in the current study.  This finding supports the use of chemiluminesence for turbulent global consumption speed measurements.

%Finally, the surface of a cone was defined to be between the rim of the burner exit and the location where $\langle c \rangle=$\num{0.5}, as illustrated in Figure \ref{fig:img_proc}(d).
%The resulting area was used as $A_{\langle c \rangle}$ when determining $S_{T,GC}$.  

\subsection{Fuel Selection}
Three fuels were considered in this study: a conventional jet-A fuel known as A2, and two experimental blends referred to here as C1 and C5. The nomenclature for the fuels is consistent with those reported previously \cite{Esclapez2017,Wang2019,Chterev2017,NJFCP2017}. 
A2 is commonly  referred to as jet-A and serves as a primary reference fuel due to its common usage in aviation.
The other two fuels have been selected to investigate potential sensitivities to fuel chemistry.
The fuels contain a bimodal blend of iso-dodecane and iso-hexadecane (C1), and a mix of iso-decane, n-decane, iso-undecane, and 1,2,4-trimethylbenzene (C5).
A2 and C5 have \SI{18}{\percent} and \SI{30}{\percent} aromatic content by mass respectively, while C1 is a pure iso-alkane blend (see Table \ref{tab:fuel_comp}).
These fuels allow sensitivities of the turbulent consumption speed to aromatic and alkane content to be identified.
Each of the fuels have similar lower heating values ($\approx$ \SI{43} {\mega\joule/\kilogram}) and densities ($\approx$ \SI{780} {\kilogram/\metre$^3$}) as indicated in Table \ref{tab:fuel_prop}.
However, these fuels have different chemical compositions, as indicated by the average molecular formulas and molar masses.
Equivalence ratio calculations were based on stoichiometry found using the average molecular formula presented in the NJFCP thermophysical properties tables \cite{NJFCP2013}.
\begin{table*}[htb!]
% \scriptsize
    \caption{Average properties for selected fuel blends, including average molecular formula, molecular weight, lower heating value (LHV), and fuel density at room temperature ($\rho$).}
\centering
    %\begin{tabular}{@{\extracolsep{\fill}}l c c c c c c c c@{}}
    \begin{tabular}{@{}c c c c c@{}}
         \toprule
         Fuel & Average Mol. Formula & Mol. Weight & LHV & $\rho$ \\
          & & [$g/mol$] & [$MJ/kg$] & $[kg/m^3]$ \\
         \midrule
         \textbf{A2} & C$_{11.4}$H$_{22.1}$ & $158.6$ & $43.0$ & $804$ \\
         \textbf{C1} & C$_{12.9}$H$_{26.8}$ & $181.9$ & $43.6$ & $782$ \\
         \textbf{C5} & C$_{9.7}$H$_{18.7}$ & $135.4$ & $42.8$ & $770$ \\
         \bottomrule
    \end{tabular}
    \label{tab:fuel_prop}
\end{table*}

\begin{table*}[htb!]
% \scriptsize
    \caption{Composition of selected fuel blends on a mass percent basis \cite{Esclapez2017}.}
\centering
    %\begin{tabular}{@{\extracolsep{\fill}}l c c c c c c c c@{}}
    \begin{tabular}{@{}c c c c c c@{}}
         \toprule
         Fuel & Aromatics & \textit{iso}-Paraffins & \textit{n}-Paraffins & Cycloparaffins& Alkenes \\
         \midrule
         \textbf{A2} & 18.66 & 29.45 & 20.03 & 31.86 & < 0.001\\
         \textbf{C1} & < 0.01 & 99.63 & <  0.001 & 0.05 & 0.32\\
         \textbf{C5} & 30.68 & 51.58 & 17.66 & 0.08 & < 0.001\\
         \bottomrule
    \end{tabular}
    \label{tab:fuel_comp}
\end{table*}

%subsection{Laminar Flame and Flame Stretch Calculations}

\section{Results}
% \subsection{Laminar Burning Parameters}

\begin{figure}[htb!] 
  \begin{center}
  \includegraphics[width=0.65\textwidth]{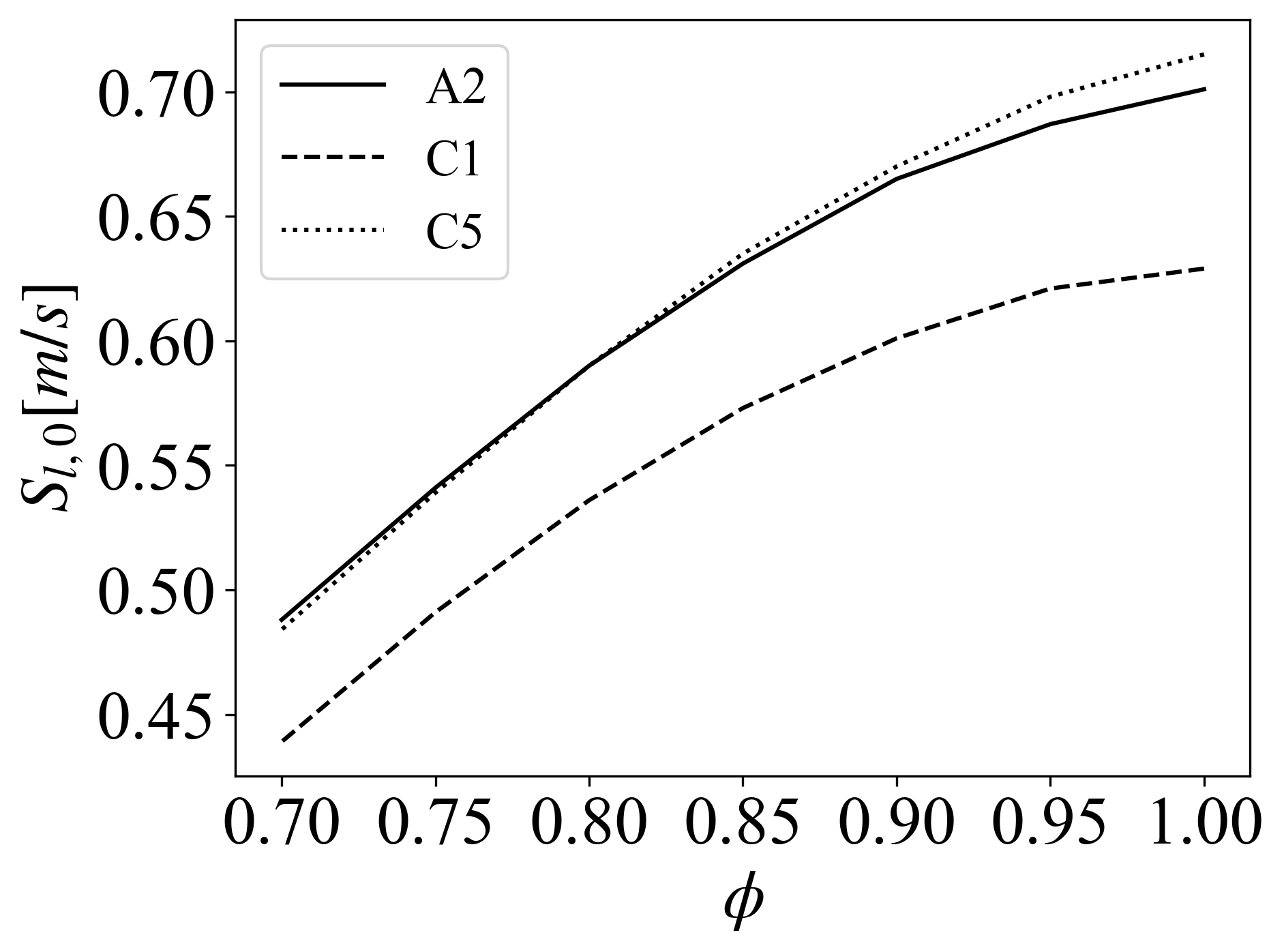}
  \caption{Zero-stretch laminar flame speeds of selected fuels at \SI{450}{\kelvin} and \SI{1}{atm}.}
  \label{laminar}
  \end{center}
\end{figure}

%\subsection{Laminar Flame and Flame Stretch Calculations}
Zero-stretch laminar flame speeds were calculated with the HyChem mechanisms \cite{WANG2018477,WANG2018502, XU2018520} in Cantera \cite{Goodwin2017}, and are reported in Figure \ref{laminar}.  The calculated flame speeds provide a reference for comparison of the turbulent consumption speed data discussed shortly.
Calculations were performed with an unburned temperature and pressure of \SI{450}{\kelvin} and \SI{101}{kPa}, respectively.
The laminar flame speed of A2 agrees within \SI{5}{\percent} of the flame speeds for C5 at all equivalence ratios.
The flame speeds for C1 are less than the other fuels, but agree within \SI{10}{\percent} of that for A2 and within \SI{15}{\percent} for C5. The observed similarities in laminar flame speeds are not surprising.  A2, C1, and C5 have similar average molecular structures and lower heating values are within \SI{2}{\percent}, as shown in Table \ref{tab:fuel_prop}.  Adiabatic flame temperatures agree within \SI{5}{\percent} for the three fuels.

\begin{figure}
    \centering
    \includegraphics[width=0.49\textwidth]{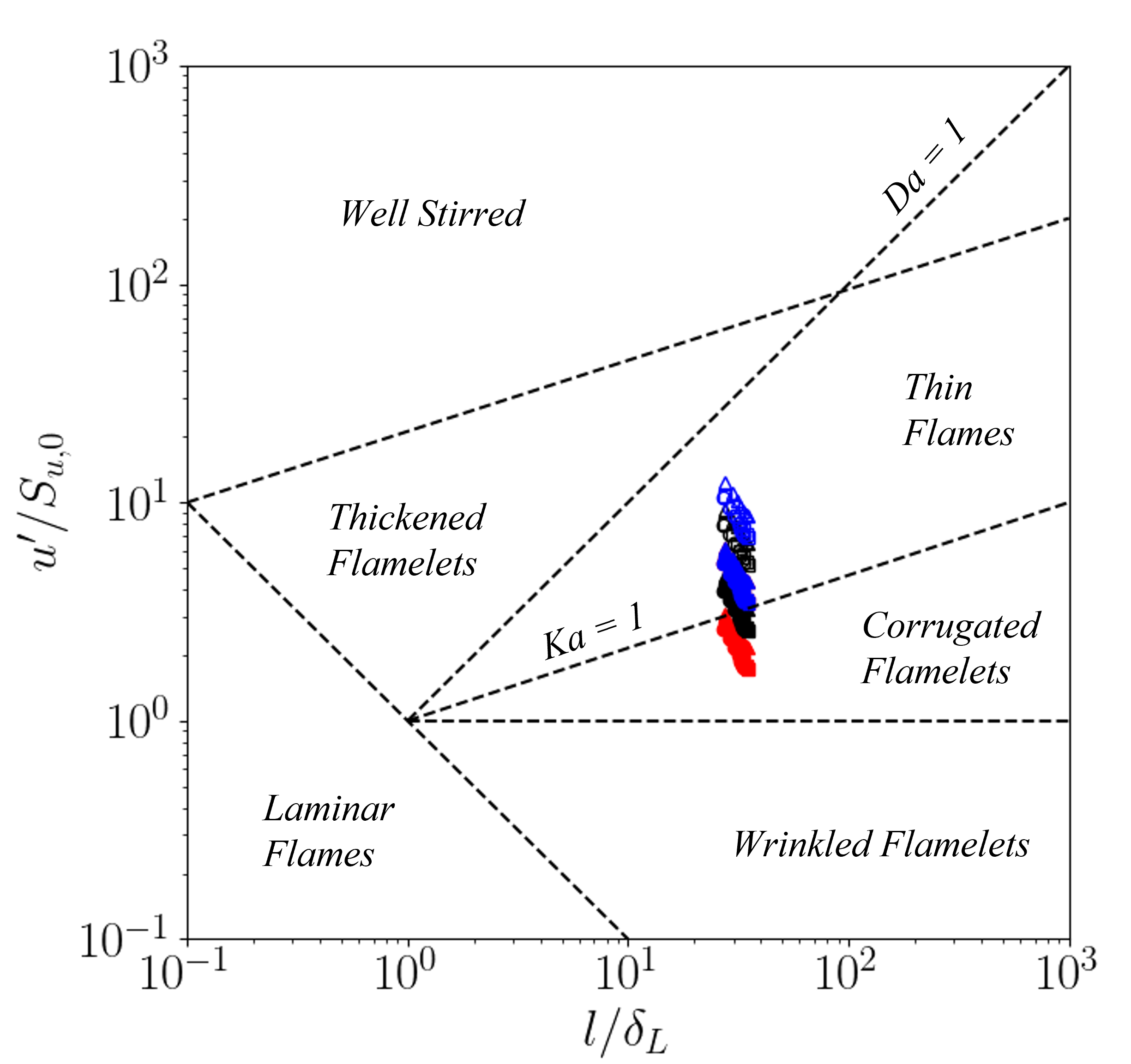}
    \caption{The orientation of the current work on a Borghi-Peterson diagram.}
    \label{fig:borghi-peterson}
\end{figure}

Using the aforementioned laminar flame speed calculations, the flames studied in this work are located on a Borghi-Peterson diagram for reference (see Figure \ref{fig:borghi-peterson}) \cite{Borghi1985OnFlames}. 
The flame thickness ($\delta_{L}$) was estimated using the location of the maximum temperature gradient obtained from laminar flame calculations \cite{Sun1999}, and the burner diameter was used as an upper bound for the integral length scale ($l$).
The turbulent flames examined in this work span the corrugated flamelet and thin flame regimes.
% Note that measurements \cite{dunn2007, zhou2017, mohammadnejad2020} and simulations \cite{wang2018DNS} of premixed turbulent flames have shown thickening of the reaction zone at high levels of turbulence.
Data presented by Mohammadnejad et al.~\cite{mohammadnejad2020} indicate that the flames examined here have a ratio of turbulent to laminar reaction zone thickness of 1-1.5.

%\subsection{Turbulent Flame Speeds}

\begin{figure*}[!ht]
  \begin{center}
    \subfigure[Low Turbulence Intensity]{\label{St_data-a}\includegraphics[width=0.49\textwidth]{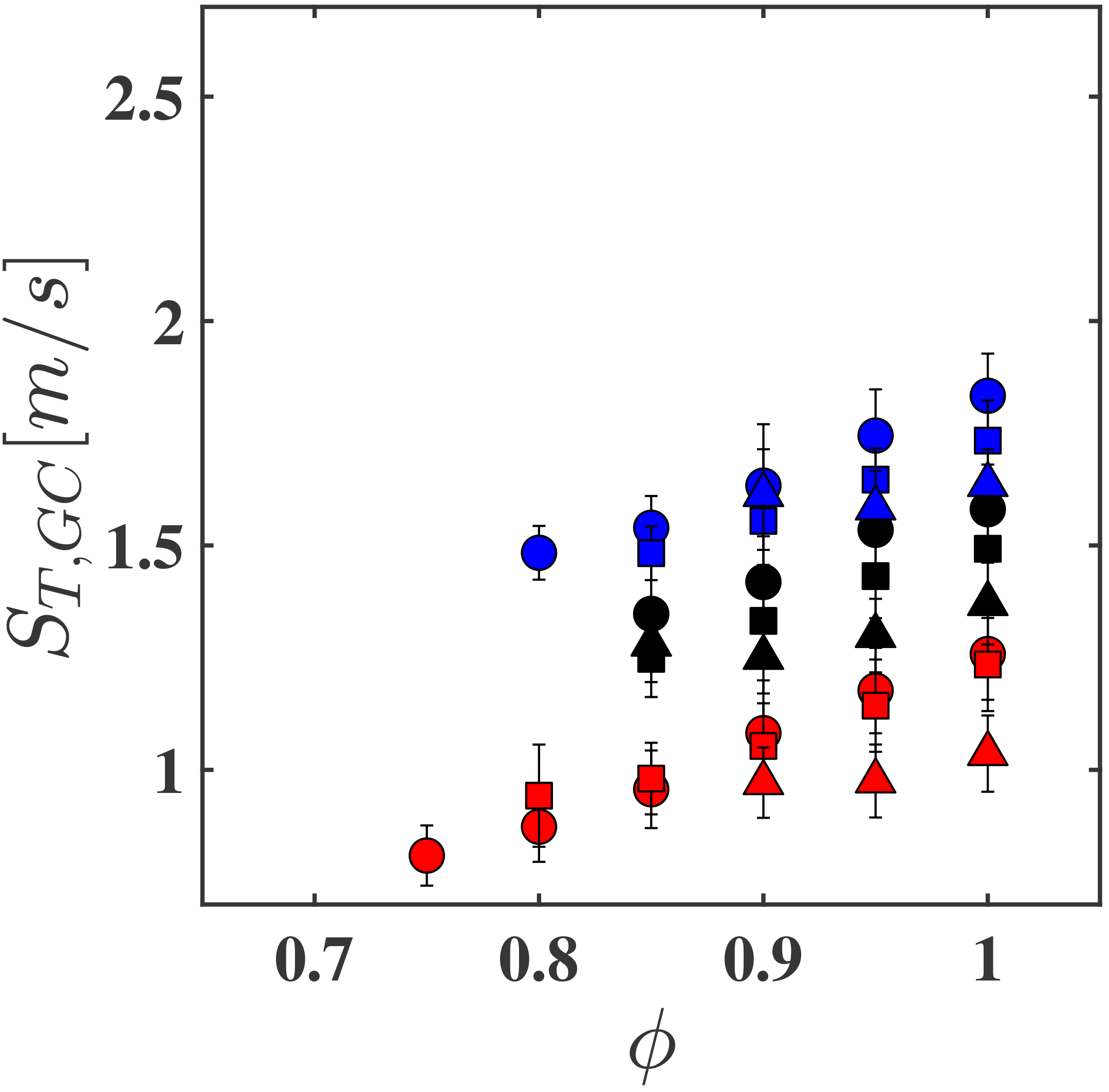}}
    \subfigure[High Turbulence Intensity]{\label{St_data-b}\includegraphics[width=0.49\textwidth]{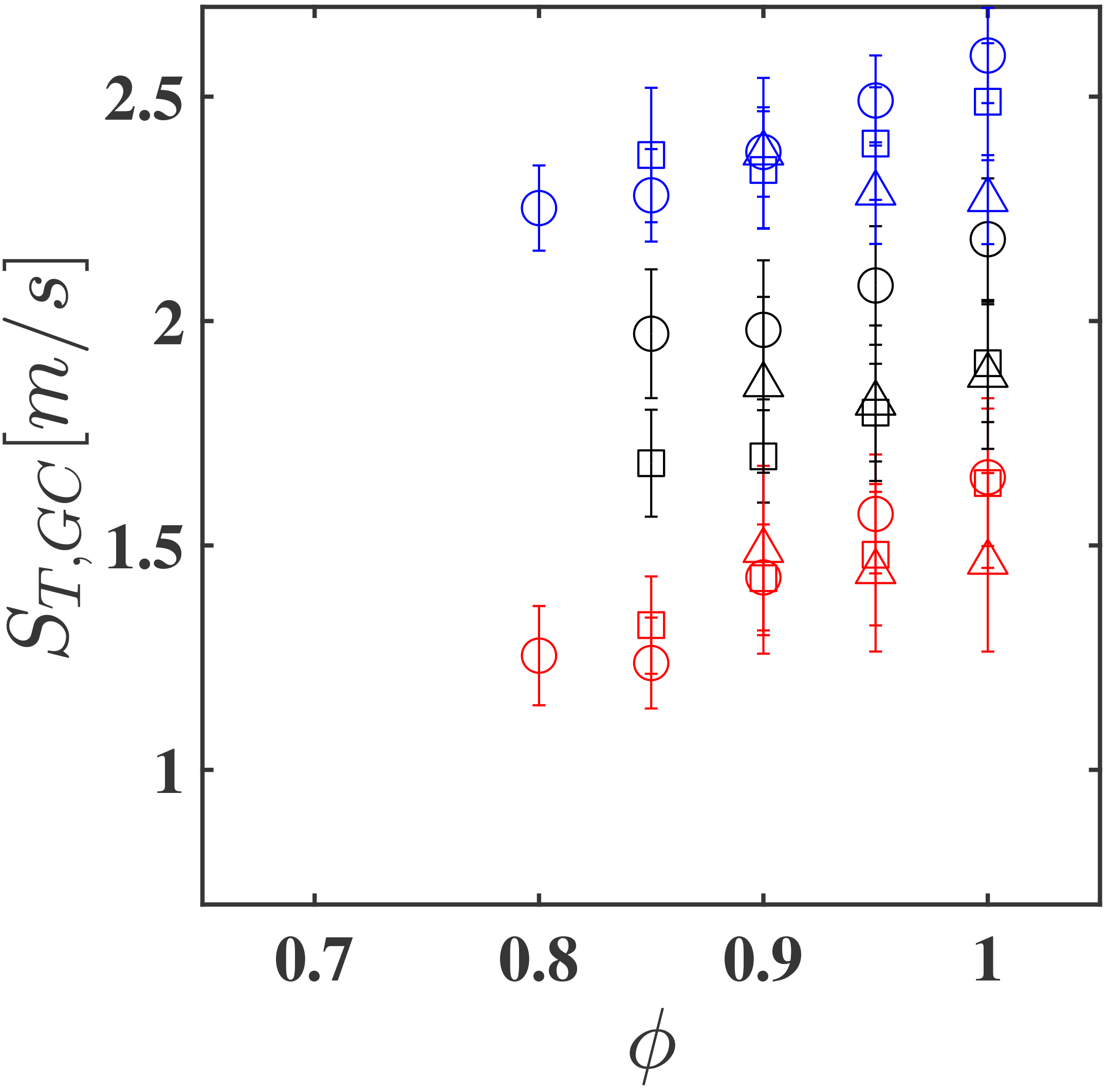}}\\
  \end{center}
  \caption{Turbulent consumption speed for all fuels and flow conditions presented relative to equivalence ratio; (a) Low turbulence intensity $I=$\SI{10}{\percent}, (b) High turbulence intensity $I=$\SI{20}{\percent}} 
  \label{St_data}
\end{figure*}

Turbulent consumption speeds for the three fuels at $Re_{D}$ of \num{5000}, \num{7500}, and \num{10000} and  turbulence intensities near \SI{10}{\percent} and \SI{20}{\percent} are presented in Figure \ref{St_data}.
Table \ref{legend} provides a summary of the symbol convention used in Fig. \ref{St_data} and in subsequent figures.
Three major trends are observed and discussed.

\begin{table*}[!ht]
% \scriptsize
    \caption{Legend of symbols representing the different operating conditions.}\label{legend}
    \begin{center}
    \begin{tabular}{@{}l c c c c c c@{}}
        \hline
         & \multicolumn{2}{c}{\textbf{A2}} & \multicolumn{2}{c}{\textbf{C1}} & \multicolumn{2}{c}{\textbf{C5}}\\
        \hline
        $Re_{D}$ & $\small I=10~\%$ & $\small I=20~\%$ & $\small I=10~\%$ & $\small I=20~\%$ & $\small I=10~\%$ & $\small I=20~\%$ \\
        \hline
        $5,000$ & \includegraphics[scale=0.4]{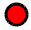} & \includegraphics[scale=0.4]{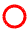} & \includegraphics[scale=0.35]{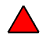} & \includegraphics[scale=0.35]{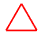} & \includegraphics[scale=0.5]{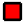} &       \includegraphics[scale=0.5]{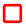} \\
        $7,500$ & \includegraphics[scale=0.4]{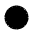} & \includegraphics[scale=0.4]{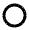} & \includegraphics[scale=0.35]{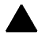} & \includegraphics[scale=0.35]{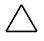} & \includegraphics[scale=0.5]{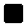} &       \includegraphics[scale=0.5]{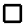} \\
        $10,000$ & \includegraphics[scale=0.4]{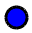} & \includegraphics[scale=0.4]{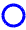} & \includegraphics[scale=0.35]{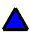} & \includegraphics[scale=0.35]{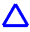} & \includegraphics[scale=0.5]{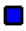} &       \includegraphics[scale=0.5]{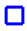} \\
        \hline
    \end{tabular}
    \end{center}
\end{table*}

First, the turbulent consumption speed for all three fuels increases with both bulk Reynolds number and turbulence intensity, consistent with similar work performed with smaller hydrocarbon fuels (i.e., syngas and methane/air ~\cite{Venkateswaran2011,Venkateswaran2015,Venkateswaran2015a}). 
The turbulent consumption speed of A2 and C5 increase with increasing equivalence ratio for both turbulence intensity conditions.
The turbulent consumption speed of A2 is within \SI{5}{\percent} of C5 for all flow conditions; similar agreement is observed for the corresponding laminar flame speeds shown in Fig. \ref{laminar}. 
This similarity is expected because of the sensitivity of the flame speed to the flame temperature, which increases as the equivalence ratio is increased \cite{Law2006}.
In contrast, for C1, there is no statistically significant sensitivity of the flame speed to equivalence ratio.
This is attributed to a heighten sensitivity of this fuel to flame stretch and the stretch induced by turbulence, as discussed later.
%In contrast, at the high turbulence intensities the global consumption speed of C1 (triangle symbols) is relatively constant with respect to changes in equivalence ratio.  
%At the lower turbulence intensity a consistent sensitivity of the global consumption speed for C1 to the equivalence ratio is not evident.  

The second trend noted is that both the A2 and C5 fuels demonstrate an increased sensitivity of the turbulent consumption speed to turbulence intensity with increasing bulk Reynolds number.
To better quantify this trend, Table \ref{Data} presents turbulent consumption speed values for A2, C1, and C5, and the percent differences in values for the two turbulence intensity levels at an equivalence ratio of $\phi =$ \num{1.0}. 
The turbulent consumption speeds of A2 and C5 both increase by \SI{24}{\percent} and \SI{30}{\percent} from low to high turbulence intensity for $Re_{D}=$ \num{5000} and $Re_{D}=$ \num{10000}, respectively.
The turbulent consumption speed of C1 increases by \SI{29}{\percent} between low and high turbulence intensity conditions at $Re_{D} =$ \num{5000}, and by \SI{27}{\percent} at $Re_{D} =$ \num{10000}.

\begin{table}[htb!]
% \scriptsize
    \caption{Representative turbulent consumption speeds at an equivalence ratio $\phi =$ \num{1.0}, for A2, C1, and C5 showing relative increase in flame speed between turbulence intensities of $I =$\SI{10}{\percent} and $I =$\SI{20}{\percent}}\label{Data}
    
    \begin{center}
    \begin{tabular}{ c c c c }
        \hline
        &  \multicolumn{2}{c}{\textbf{$S_{T,GC}~[m/s]$}} & \\
        \hline
        \multicolumn{4}{c}{\textbf{A2}} \\
        $Re_{D}$ & $I=10\ \%$ & $I=20\ \%$ & \textbf{Increase} \\
        \hline
        $5,000$ & $1.26$ & $1.65$ & $\mathbf{24\ \%}$ \\
        $7,500$ & $1.58$ & $2.18$ & $\mathbf{28\ \%}$ \\
        $10,000$ & $1.83$ & $2.59$ & $\mathbf{30\ \%}$ \\
        \hline
        \multicolumn{4}{c}{\textbf{C1}} \\
        \hline
        $5,000$ & $1.04$ & $1.46$ & $\mathbf{29\ \%}$ \\
        $7,500$ & $1.37$ & $1.88$ & $\mathbf{27\ \%}$ \\ 
        $10,000$ & $1.64$ & $2.27$ & $\mathbf{27\ \%}$  \\
        \hline
        \multicolumn{4}{c}{\textbf{C5}} \\
        \hline
        $5,000$ & $1.23$ & $1.64$ & $\mathbf{24\ \%}$ \\
        $7,500$ & $1.49$ & $1.91$ & $\mathbf{22\ \%}$ \\ 
        $10,000$ & $1.73$ & $2.49$ & $\mathbf{30\ \%}$\\
        \hline
    \end{tabular}
    \end{center}
\end{table}

Third, it is noted that the leanest equivalence ratio for which a flame speed is reported for each $Re_{D}$, and turbulence intensity is different for each fuel.  
As the equivalence ratio is decreased sufficiently, all fuels exhibit a behavior in which the flame tip opens (i.e., tip quenching).  The tip quenching phenomena is consistent with the work of others burning premixed large hydrocarbon fuels, as discussed shortly \cite{carbone2017comparative}.
Of interest here is that tip opening occurs at different conditions for different fuels (see Table \ref{limits}), as indicated by the lowest equivalence ratio for which data is reported.
Opening of the flame tip was confirmed by visual assessment of individual UV images. When the flame tip is open the conical geometry of the flame is disrupted and the measurement technique is no longer valid because fuel can escape through the  front. Values associated with tip-quenching are not reported.

A2 has the lowest $\phi$ without tip quenching (i.e., $\phi=$ \num{0.80}) whereas C5 displays closed flame tips to $\phi=$ \num{0.85} and C1 has closed flame tips to $\phi=$ \num{0.90}, as shown in Table \ref{limits}.
The lean limit of A2 and C5 extends to $\phi=$ \num{0.75} and $\phi=$  \num{0.8}, respectively, for $Re_{D}=$ \num{5000} and lower turbulence intensity.
The reduced tip stability of C1 highlights a sensitivity of tip quenching to fuel chemistry, and is consistent with the reduced reactivity of C1.
Carbone and coworkers \cite{carbone2017comparative} attributed quenching near the tip to the higher strain rates and entrainment of the surrounding cold air into the shear layer.
A sensitivity of tip quenching to the reactivity of the fuel was noted in their work.  

%When laminar Bunsen flames are stretched beyond a critical value, they can break open at the tip \cite{Law2006}.
%Hui et al.~\cite{Hui2012,Hui2013}, Kumar et al.~\cite{Kumar2007,Kumar2011}, and others~\cite{Ji2012,Sarathy2013} have observed that flame stretch rate increases with increased alkane content.
%C1 is \SI{100}{\percent} iso-alkanes and has no aromatic content but is the least stable of the three fuels; conversely, A2 and C5 are \SI{18}{\percent} and \SI{30}{\percent} aromatic content by mass, respectively.
%Moreover, flame stretch is a thermal loss phenomena which reduces the flame temperature and can result in flame extinction.
%Thus, it is plausible that the high iso-alkane content of C1 makes it more sensitive to turbulence induced stretch effects, ultimately decreasing the range over which tip quenching was absent.

\begin{table*}[tbh!]
% \scriptsize
    \caption{Approximate leanest equivalence ratio for which no tip quenching was observed for A2, C1, and C5 flames relative to Reynolds number and turbulence intensity.}\label{limits}
    %\vspace{-1.5em}
    \begin{center}
    \begin{tabular}{ c c c c c c c }
        \hline
         & \multicolumn{2}{c}{\textbf{A2}} & \multicolumn{2}{c}{\textbf{C1}} & \multicolumn{2}{c}{\textbf{C5}}\\
        \hline
        $Re_{D}$ & $I = 10\%$ & $I = 20\%$ & $I = 10\%$ & $I = 20\%$ & $I = 10\%$ & $I = 20\%$ \\
        \hline
        $5,000$ & $0.75$ & $0.80$ & $0.90$ & $0.90$ & $0.80$ & $0.85$ \\
        $7,500$ & $0.85$ & $0.85$ & $0.85$ & $0.90$ & $0.85$ & $0.85$ \\
        $10,000$ & $0.80$ & $0.80$ & $0.90$ & $0.90$ & $0.85$ & $0.85$ \\
        \hline
    \end{tabular}
    \end{center}
\end{table*}

\begin{figure*}[!ht]
  \begin{center}
    \subfigure[Turbulent Flame (I = \SI{10}{\percent})]{\label{St_Tad-b}\includegraphics[width=0.49\textwidth]{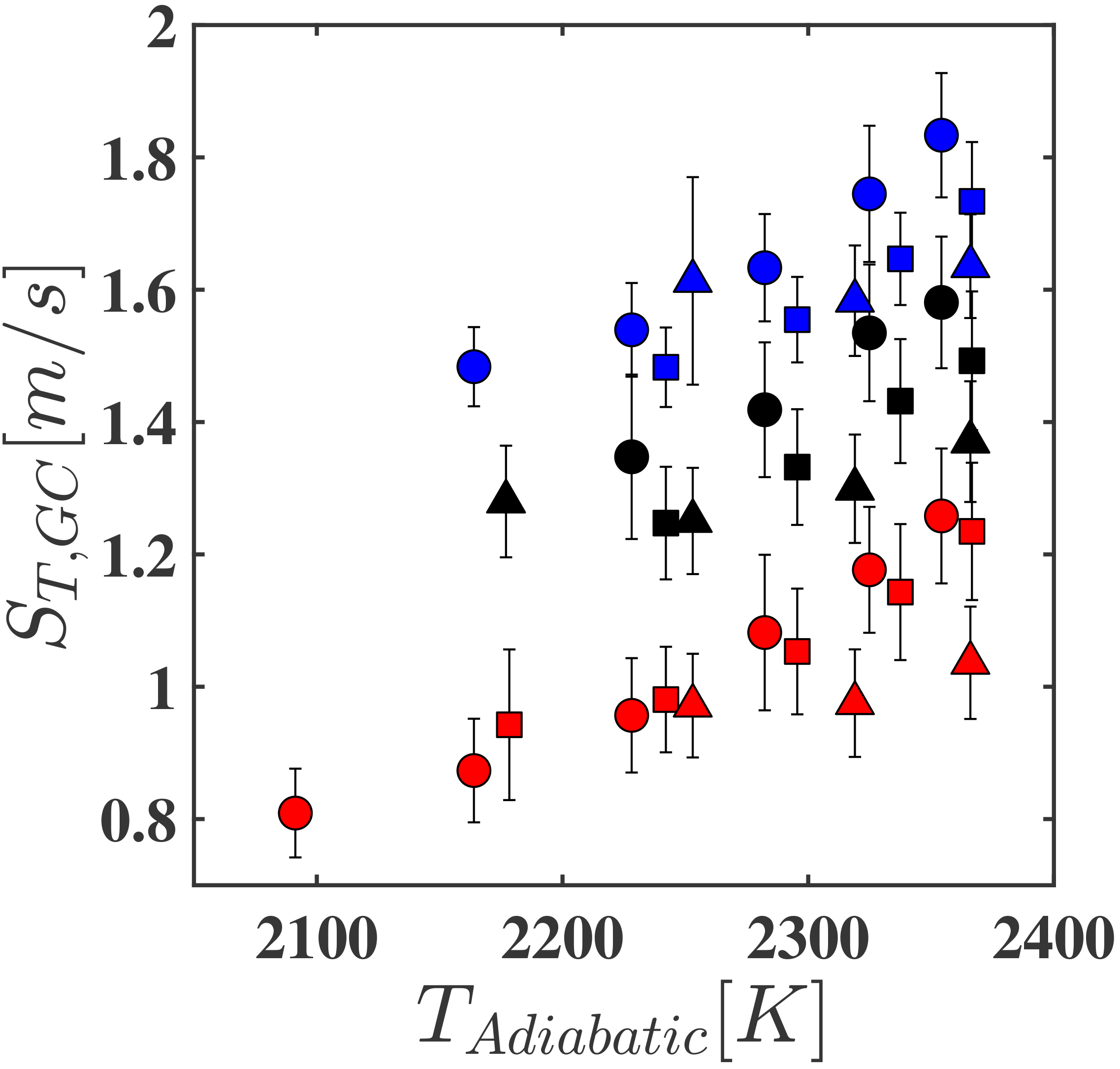}}
    \subfigure[Laminar Flame]{\label{St_Tad-a}\includegraphics[width=0.49\textwidth]{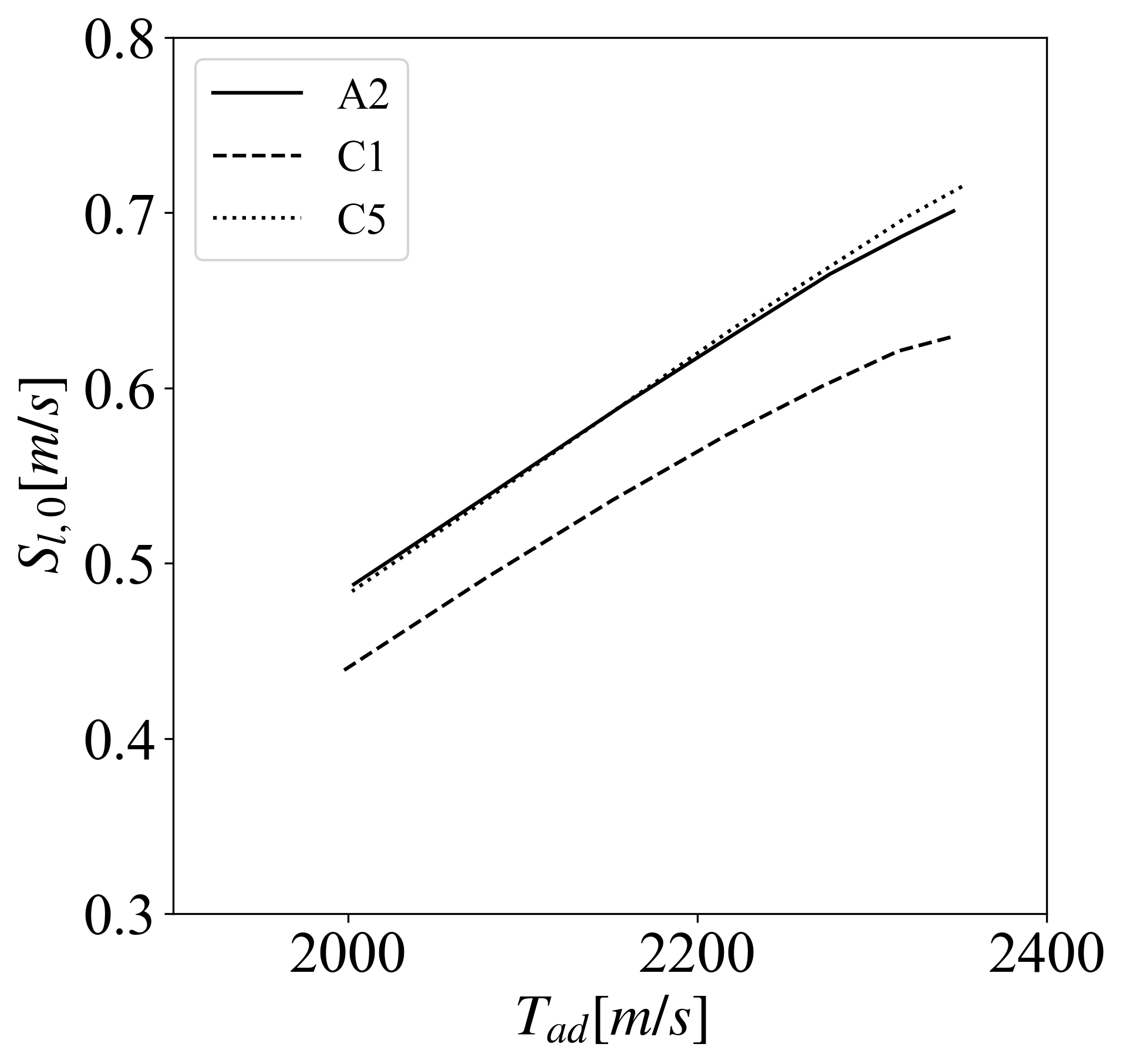}}\\
  \end{center}
  \caption{(a) Turbulent consumption speed versus adiabatic flame temperature for all fuels at low turbulence intensity of $I=$\SI{10}{\percent}.  The corresponding  laminar flame speeds plotted relative to $T_{ad}$ are shown in panel (b) for the same fuels.  See Table \ref{legend} for a summary of the symbol conventions.} 
  \label{St_Tad}
\end{figure*}

%\subsection{Flame Temperature Analysis}
The turbulent consumption speeds for the three fuels at an estimated turbulence intensity of \SI{10}{\percent} are presented relative to adiabatic flame temperatures in Figure \ref{St_Tad}(a) to allow temperature effects on flame speeds to be identified.
The corresponding laminar flame speeds are presented in Figure \ref{St_Tad}(b). 
The adiabatic flame temperatures were calculated using the NJFCP thermophysical property tables \cite{NJFCP2013}.
Similar trends to those in Fig. \ref{St_Tad}(a) were observed for the higher turbulence intensity conditions, and are not reported for brevity.

Both the laminar and turbulent consumption speeds increase with increasing adiabatic flame temperature up to an equivalence ratio of unity, as expected \cite{Law2006}.  
A2 and C5 have statistically similar flame speeds while C1 typically has lower values, even for similar adiabatic flame temperatures.  
This observation is significant because it indicates that the different burning behavior of C1 is caused by processes other than changes in flame temperatures.  

%Moreover, there is no observable difference in the sensitivities of the laminar and turbulent flame speeds to the adiabatic flame temperature.
%This suggests there is no difference in the sensitivity of the global consumption speed to adiabatic flame temperature relative to the laminar flame speed.

\begin{figure*}[htb!]
    \centering
    \includegraphics[width=0.6\textwidth]{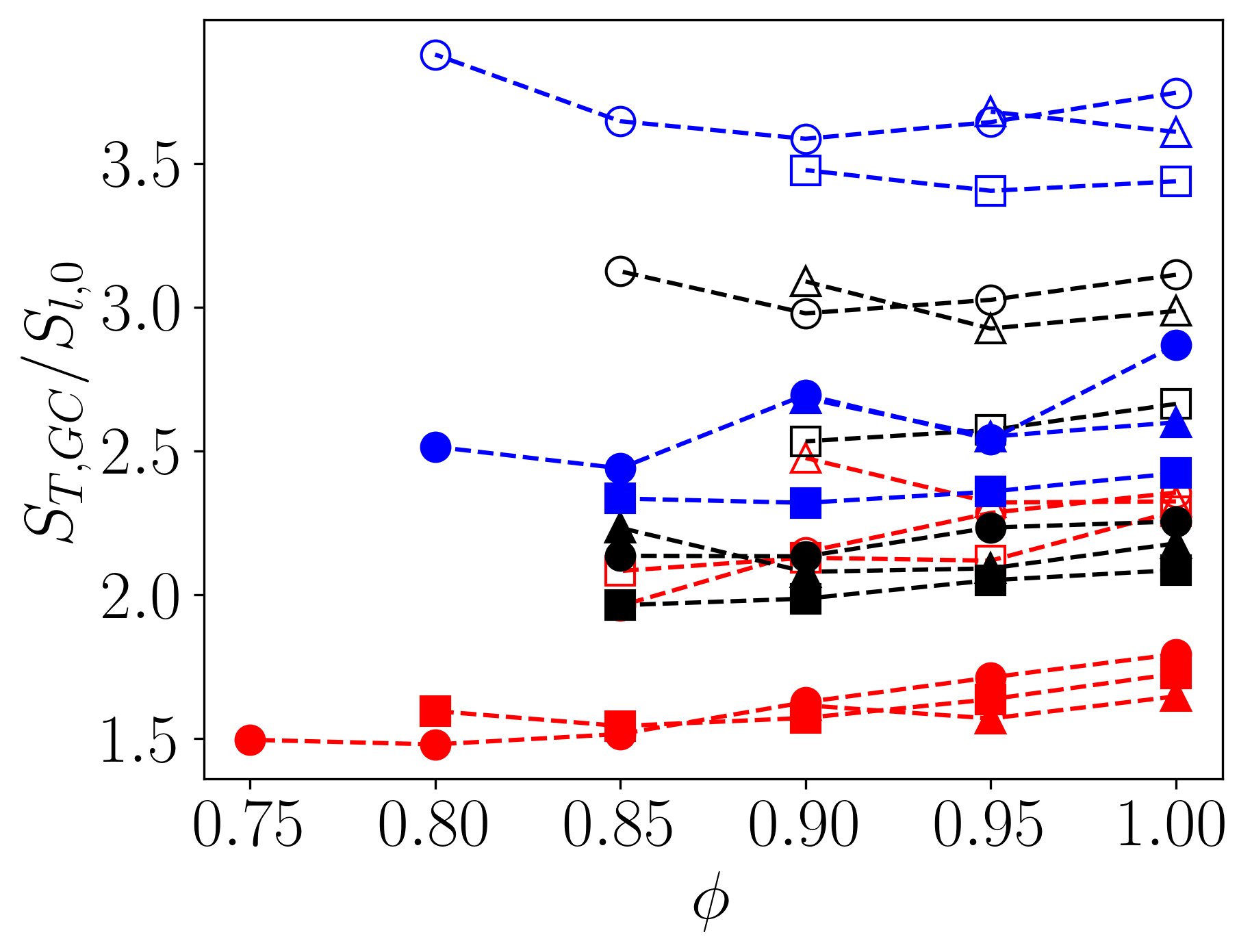}
 \caption{Turbulent consumption speed normalized by unstretched laminar flame speed for varying equivalence ratios. Table \ref{legend} provides a summary of the symbol conventions.} 
  \label{normalized_flame_speeds}
\end{figure*}

%\subsection{Flame Stretch Analysis}
Figure \ref{normalized_flame_speeds} shows the turbulent consumption speed values normalized by the corresponding zero-stretch laminar flame speed.
These results are shown to help evaluate sensitivities of the turbulent consumption speeds to the various fuels that are not captured by changes in laminar flame speeds.
Turbulent consumption flame speeds are roughly 1.5 to 4.5 times larger than corresponding laminar flame speeds, depending on the fuel and flow conditions.
For equivalent flow conditions, normalized flame speeds are highest for C1 and similar for A2 and C5 fuels.
This observation illustrates a greater sensitivity of C1 flames to the presence of turbulence than the other fuels.
Irrespective of the fuel, normalized flames speeds are roughly constant as the equivalence ratio is varied (i.e., agree within $I=$\SI{10}{\percent}).
It is noted that at the lowest equivalence ratios, normalized values tend to increase.
It is plausible that the increased relative flame speeds at the lowest equivalence ratios results from heightened stretch sensitives (as discussed later), or fuel breaking through the flame front as the flames approach tip-quenched conditions.
The latter process would bias the measurements higher based on equation 1 because not all of the fuel would be consumed. 

\begin{figure*}[!ht]
  \begin{center}
    
    \subfigure[Sensitivity to $u_{rms}'$]{\label{U_urms_norm-a}\includegraphics[width=0.49\textwidth]{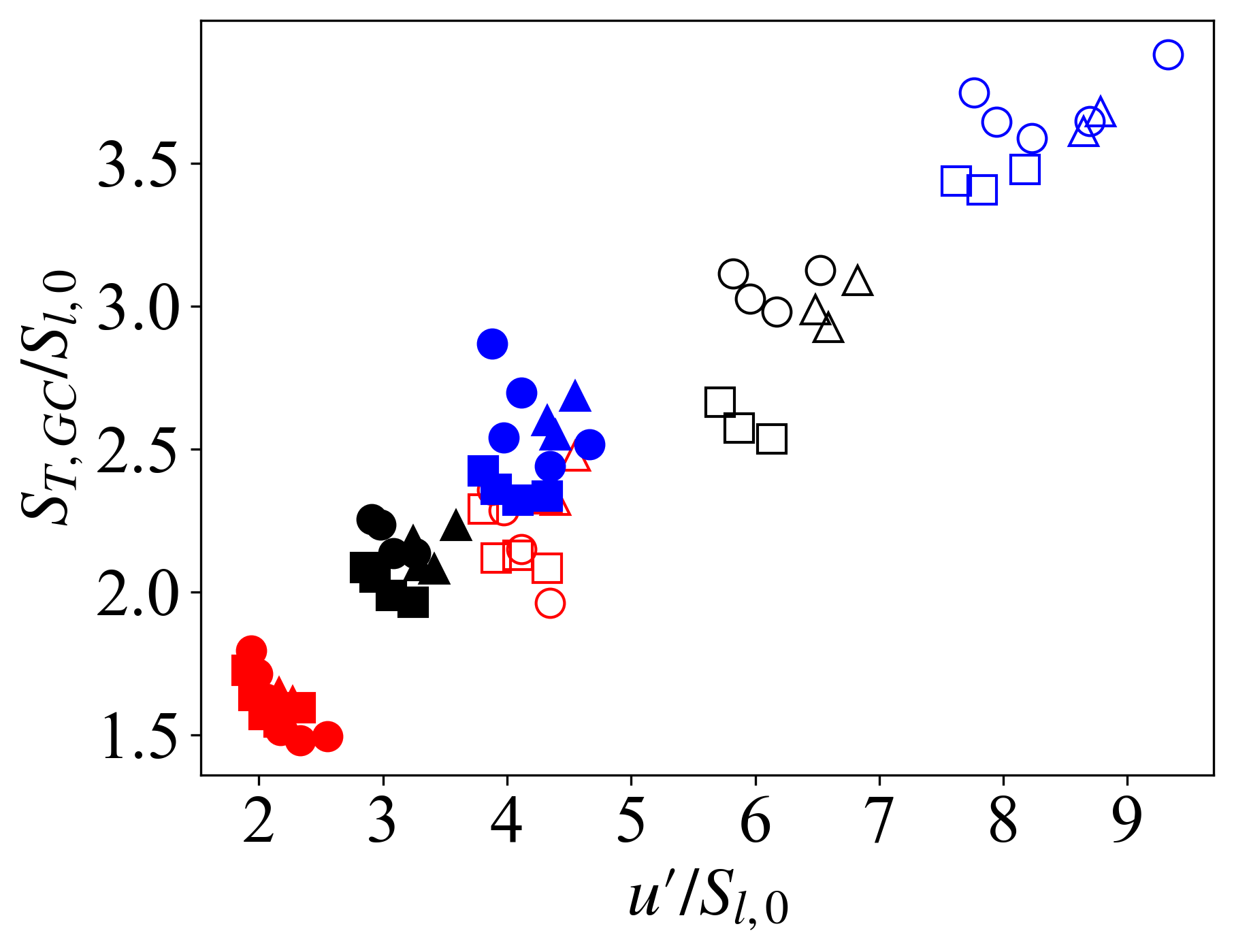}}
    \subfigure[Sensitivity to $U_0$]{\label{U_urms_norm-b}\includegraphics[width=0.49\textwidth]{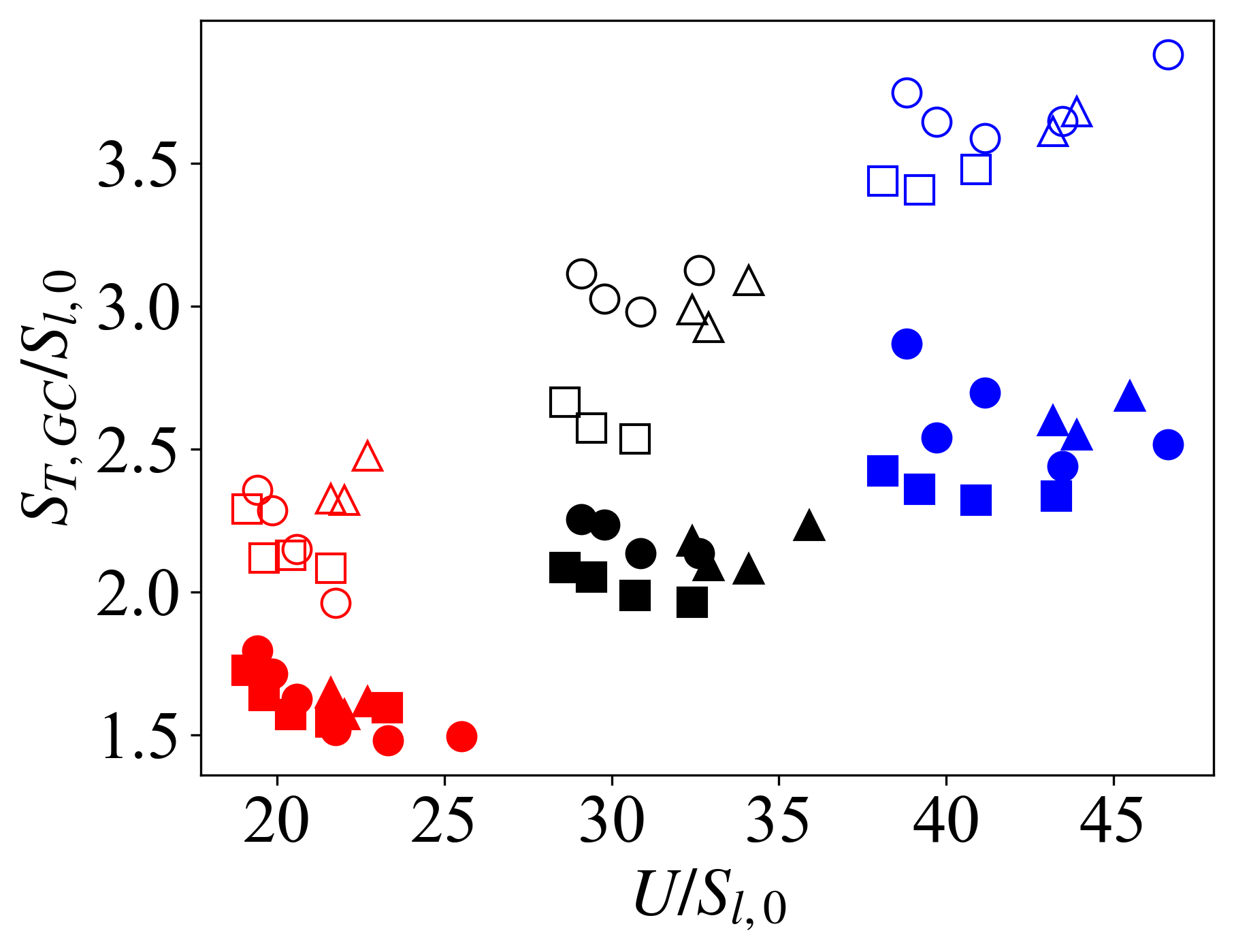}}\\
  
  \end{center}
  
  \caption{Global consumption speed relative to the rms of the velocity fluctuations (panel a) and the bulk flow velocity normalized by zero-stretch laminar flame speed (panel b).  The equivalence ratio increases from left to right for each data set.  Table \ref{legend} provides a summary of the symbol conventions.} 
  \label{U_urms_norm}

\end{figure*}

Figure \ref{U_urms_norm} presents the normalized flame speeds relative to both the estimated root mean square (rms) of the velocity fluctuations [Figure \ref{U_urms_norm-a}] and the bulk fluid velocity [Figure \ref{U_urms_norm-b}].
This normalization is performed to help isolate the effects of local and bulk flow velocities on the global consumption speed.
Similar analyses have been performed previously for small hydrocarbon fuels \cite{Venkateswaran2011,Venkateswaran2015,Kobayashi2002,Kobayashi2005,Daniele2012,Daniele2011TurbulentConditions}.
Note that equivalence ratios increase from left to right for each cluster of data shown in Figure \ref{U_urms_norm}.
Uncertainty bars have been omitted for clarity; and are on the order of \SI{10}{\percent} or less of the plotted value on both axes.

Three key trends are noted in Figure \ref{U_urms_norm}.
First, the turbulent consumption speed increases proportionally for increasing normalized $u_{rms}'$, as shown in panel (a).  It is noted that the normalized flame speeds have similar values (i.e., agree within roughly 25\%) to those corresponding to \ce{H_{2}}/CO mixtures \cite{venkateswaran2015scaling} and \ce{CH_{4}} \cite{Daniele2011TurbulentConditions} for similar normalized turbulence fluctuations.
This observation suggests that the turbulent consumption speeds of jet fuels are sensitive to hydrodynamic stretch.
Second, the turbulent consumption speed is proportional to $U_{0}$, as shown in panel (b), with different proportional relationships observed depending on the turbulence intensity.
A greater sensitivity is observed at higher turbulence intensity conditions (i.e., steeper slope).  
In contrast, the sensitivity of consumption speeds to $u_{rms}'$ has a similar relationship for high and low turbulence intensities.
The third observation is that the trends in the normalized values tend to collapse for the various fuels.    
%Both trends suggest both a sensitivity of the global consumption speed to global flame curvature, and a strong coupling of hydrodynamic and curvature based stretch effects.
These observations are consistent with similar work examining smaller hydrocarbon fuels and suggest that the current understanding of the influence of turbulence on combustion of gaseous fuels may apply, in general, to large hydrocarbon fuels as well~\cite{Venkateswaran2011,Venkateswaran2015a,Kobayashi2005}.  

Markstein numbers (i.e., ${L}_{b}/l_{f}$) were calculated for different equivalence ratios to evaluate the potential role of stretch sensitivities in causing C1 flames to have different tip quenching and turbulent flame speed behaviors than C5 and A2.
Here, ${L}_{b}$ is the burned Markstein length, which is an indication of the stretch sensitivity of a fuel, with larger values corresponding to greater sensitivity.
The parameter $l_{f}$ is the laminar flame thickness, which was calculated from the temperature gradient within laminar flame simulations.

${L}_{b}$ values were determined using the expression, from Sun et al.~\cite{Sun1999}: 
\begin{equation}
    S_{L,b} = S_{L,b0}-{L}_{b}\kappa~.
\end{equation}
Here, $S_{L,b}$ is the burned laminar flame speed evaluated at the strain rate $k$, and $S_{\ce{L,b0}}$ is the unstretched burned laminar flame speed.
Flame simulations were performed using Cantera \cite{Goodwin2017} and the HyChem mechanisms for jet-A \cite{WANG2018502, XU2018520}, C5\footnote{Hai Wang, Personal communication, 2016,  the method with which the model is derived can be found in \cite{WANG2018502, XU2018520}}, and C1 \cite{WANG2018477}. 
$S_{\ce{L,b}}$ was evaluated at the location of peak heat release in the strained laminar flame, and $S_{\ce{L,b0}}$ was determined by extrapolating $S_{\ce{L,b}}$ to zero-stretch.

\begin{figure*}[!ht]
  \begin{center}
  %  \subfigure[Global consumption speed relative to the global stretch rate for an equivalence ratio of, $\phi=$\num{1.0}.]{\label{stretch-a}\includegraphics[width=0.49\textwidth]{St_kappa.pdf}}
    \includegraphics[width=0.5\textwidth]{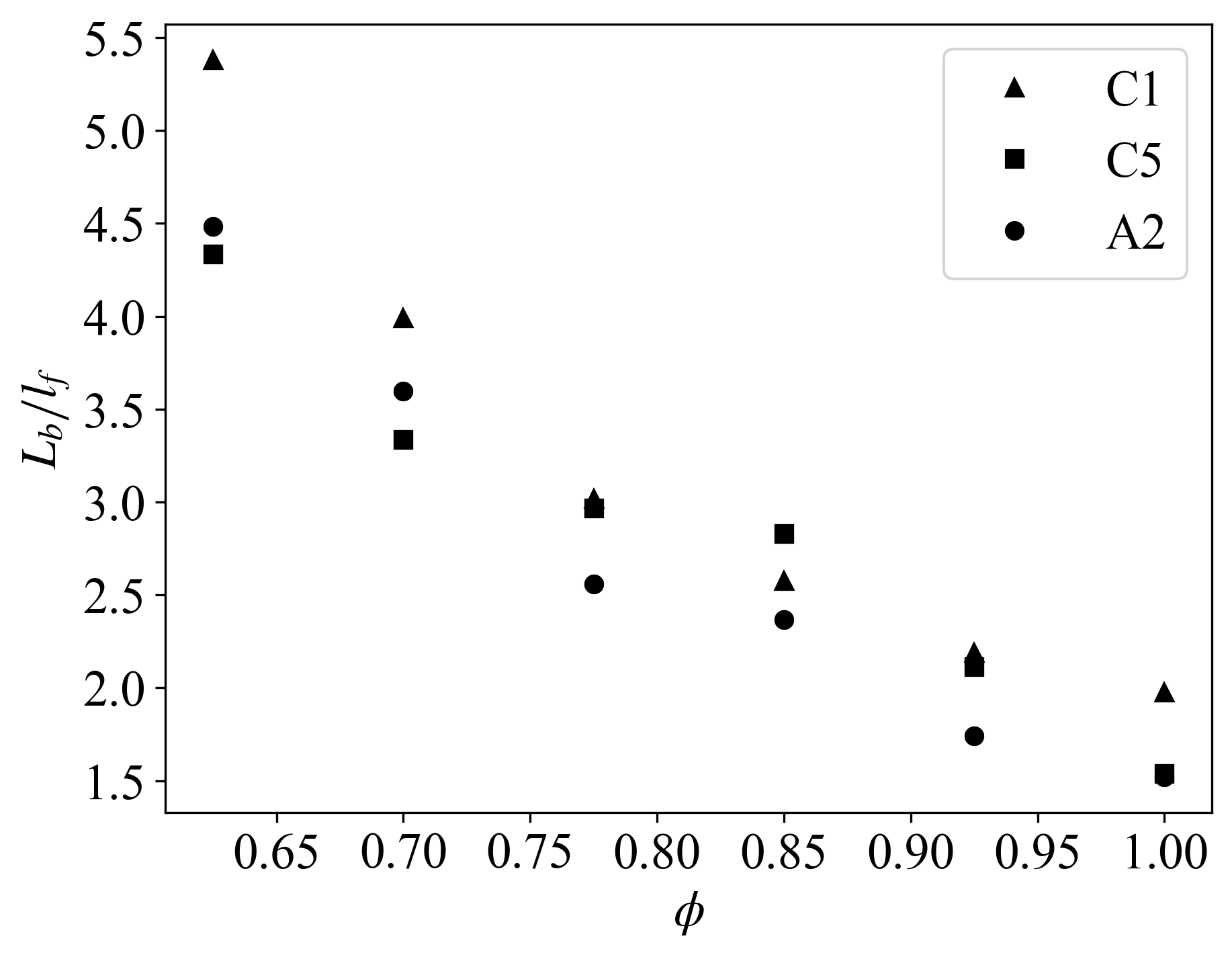}\\
  \end{center}
  \caption{Burned Markstein numbers (${L}_{b}/l_{f}$) calculated for mixtures with $ 0.625 \leq \phi \leq 1.0 $.} 
  \label{Lb}
\end{figure*}

The Markstein numbers for A2, C1 and C5 are presented in Fig. \ref{Lb}. 
At stoichiometric conditions, the Markstein number for C1 is the largest.
As the equivalence ratio decreases the magnitude of the Markstein number increases for the three fuels.
The Markstein numbers for C1 and C5 become more similar for the intermediate equivalence ratios evaluated.
At the lowest equivalence ratio the Markstein numbers for C5 and A2 approach each other.
The nonlinear relationship between changes in equivalence ratio and Markstein numbers for the various fuels is consistent with the behavior of ${L}_{b}$ of other fuels (e.g., dimethyl ether-air~\cite{yu2015effects}).
These results suggest that C1 is the most sensitive to flame stretch, in general, while A2 is the least sensitive to flame stretch across the equivalence ratios of interest in this study.

The calculated stretch sensitivities illustrated in Figure \ref{Lb} are consistent with the observed turbulent combustion behavior of the various fuels.
Recall that tip opening for C1 flames occurred at higher equivalence ratios (see Table 4) than A2 or C5 flames, consistent with the greater stretch sensitivity of C1 (in general).
Moreover, C5 flames experienced tip quenching at higher equivalence ratios than those for A2, also consistent with the larger stretch sensitivity of C5 as compared to A2.
Note that Bunsen flames break open at the tip when sufficient stretch occurs \cite{Law2006}.

Trends in turbulent consumption speeds are also consistent with trends in Markstein numbers. 
Specifically, C1 flames have the highest ratio of turbulent to laminar flame speeds of the fuels evaluated. 
It is likely that the greater stretch sensitivity of C1 fuels causes the larger increases in flame speeds when turbulence is present.

% Moreover, these trends support the notion that flame stretch alters the flame area and structure, changing the global consumption speed.
% Further investigation of the impact of flame tip opening at the stability limits would validate these theories.

The results from this work align with what has been reported previously regarding the role of different classes of hydrocarbon fuels in influencing stretch sensitivities.
Specifically, that stretch sensitivity increases with increased alkane content~\cite{Hui2012,Hui2013,Kumar2007,Kumar2011,Ji2012,Sarathy2013}.  
Hui et al.~\cite{Hui2012,Hui2013} and Kumar et al.~\cite{Kumar2007,Kumar2011} highlighted that fuels with significant aromatic content can have a weak stretch sensitivity such that the zero-stretch and max-stretch laminar flame speeds are nearly equal ($S_{L,max}/S_{L,0}\approx1$).
Alternatively, Venkateswaran et al.~\cite{Venkateswaran2015} demonstrated that alkane fuels such as \ce{CH4} have a higher stretch sensitivity, where the max-stretch laminar flame speed is much larger than the zero-stretch laminar flame speed ($S_{L,max}/S_{L,0}\gg1$).
Fuels which are highly sensitive to flame stretch, such as fuels with high alkane content, will exhibit a higher stretch rate resulting in greater thermal losses in the flame and thus a reduced flame tip stability (i.e., as evident by tip quenching).
As previously stated, C1 is strictly a blend of iso-dodecane and iso-hexadecane and has no aromatic content; A2 and C5 are $\approx$ \SI{18}{\percent} and $\approx$ \SI{30}{\percent} by mass aromatic species, respectively.
Thus the results from this work are consistent with previous studies that have considered the role of fuel composition on stretch sensitivities, among other factors.

\section{Summary and Conclusions}
Turbulent consumption speeds for large hydrocarbon fuels relevant to aviation are reported in this work.
For constant $Re_D$ and turbulence intensity, A2 (i.e., jet-A) has the highest turbulent flame speeds and remains stable (without tip quenching) at lower $\phi$ than C1 and C5. In contrast C1, which contains no aromatics, has the slowest turbulent flame speeds and exhibits tip quenching at higher $\phi$ than A2 and C5.
C1 was the most sensitive to the influence of turbulence, as evident by this fuel having the largest ratio of turbulent to laminar flame speeds.
Finally, C5 has a tip stability limit between that of the other two fuels, but similar global consumption speeds to A2 (i.e., generally agrees within 6 \%). 

%Moreover, this fuel sensitivity is evident from difference sensitivities to tip openings of the flames for the three fuels.
%Moreover, the fuel sensitivity is evident in the onset of tip opening at different equivalence ratios for each of the three fuels.
This work shows that the turbulent combustion behavior (i.e., turbulent flame speeds and tip stability) of multi-component large hydrocarbon fuels can be sensitive to the chemical class of its components, even for fuels with similar heat releases (within 2 \%), laminar flame speeds (within 5-15 \%), and adiabatic flame temperatures.
This conclusion is attributed to stretch sensitivities of the fuels, among other factors.  The results from the current work indicate that caution may be needed when using alternative or surrogate fuels to replicate conventional fuels, especially if the alternative fuels are missing chemical classes of fuels that influence stretch sensitivities.
In this study, the absence of aromatics for the C1 fuel seems to correlate with the differences in flame behavior.
    %The normalized global consumption speed of C1 is consistently \SI{10}{\percent} higher for a given stretch rate than A2 and C5.
    %Aromatic fuels have a weak sensitivity to flame stretch while alkane fuels have a stronger sensitivity to flame stretch \cite{Hui2012,Hui2013,Kumar2007,Kumar2011,Venkateswaran2015}.
    % Both A2 and C5 are approximately \SI{18}{\percent} and \SI{30}{\percent} by mass aromatic fuels, respectively, while C1 has no aromatic content.
    % Thus, it is plausible that aromatic content decreases a flame's sensitivity to stretch.

    % Flame stretch constitutes an energy loss and lowers the flame temperature \cite{Law2006}.
    % This reduced temperature weakens the flame structure at the tip and can cause it to break open in high stretch conditions.  

\begin{acknowledgement}
This work was funded by the US Federal Aviation Administration (FAA) Office of Environment and Energy as a part of ASCENT Project 27B FAA Award Number: 13-C-AJFE-OSU-02 and in part from the National Science Foundation under Grant No.~1314109-DGE.

The authors would also like to acknowledge the contributions of: Tim Lieuwen at The Georgia Institute of Technology for sharing the design of their gaseous, premixed, turbulent Bunsen burner; Tim Edwards at the Air Force Research Laboratory for supplying the liquid jet fuels studied; Scott Stouffer at the University of Dayton for sharing the premixed vaporizer design, and Hai Wang and Rui Xu from Stanford University for the chemical kinetic models used in this study.  Sampath Adusumilli provided guidance about performing the stretch sensitivity calculations.
\end{acknowledgement}

% \section*{References}
%
\bibliography{mendeley}
\bibstyle{achemso}

\end{document}